\documentclass[journal]{IEEEtran}
\usepackage{graphicx}
\usepackage{graphics} 
\usepackage{epsfig} 
\usepackage{amssymb}  
\usepackage{graphics} 
\usepackage{bm,upgreek}
\usepackage{amsmath}
\usepackage{multirow}
\usepackage[ruled,vlined]{algorithm2e}
\usepackage[compatible]{algpseudocode} 
\usepackage{cite}
\usepackage[dvipsnames]{xcolor}
\usepackage{tikz}
\usepackage{url}
\usepackage{cite}
\usepackage{lipsum}
\usepackage{makecell}

\IEEEoverridecommandlockouts

\makeatother

\title{\LARGE \bf Effective Multi-Agent Deep Reinforcement Learning Control with \\Relative Entropy Regularization}

\author{Chenyang Miao$^{1,2}$, Yunduan Cui$^{2,*}$, Huiyun Li$^{2}$ and Xinyu Wu$^{2}$
\thanks{This research is supported in part by the National Natural Science Foundation of China under Grants 62103403; in part by Guangdong Basic and Applied Basic Research Foundation under Grant 2020B515130004.}
\thanks{$^{1}$ University of Chinese Academy of Sciences, China}
\thanks{$^{2}$ Guangdong-Hong Kong-Macao Joint Laboratory of Human-Machine Intelligence-Synergy Systems, Shenzhen Institute of Advanced Technology, Chinese Academy of Sciences, Shenzhen, China.}
\thanks{$^*$ Corresponding author: Yunduan Cui (e-mail: cuiyunduan@gmail.com)}
}

\begin{document}

\maketitle

\begin{abstract}
In this paper, a novel Multi-agent Reinforcement Learning (MARL) approach,  Multi-Agent Continuous Dynamic Policy Gradient (MACDPP) was proposed to tackle the issues of limited capability and sample efficiency in various scenarios controlled by multiple agents.
It alleviates the inconsistency of multiple agents' policy updates by introducing the relative entropy regularization to the Centralized Training with Decentralized Execution (CTDE) framework with the Actor-Critic (AC) structure.
Evaluated by multi-agent cooperation and competition tasks and traditional control tasks including OpenAI benchmarks and robot arm manipulation, MACDPP demonstrates significant superiority in learning capability and sample efficiency compared with both related multi-agent and widely implemented signal-agent baselines and therefore expands the potential of MARL in effectively learning challenging control scenarios.
The open source code of MACDPP is available at \url{https://github.com/AdrienLin1/MACDPP}. 
\end{abstract}


%

%
%
%
%

\section{Introduction}\label{S1}

\IEEEPARstart{G}{uided} by task-related reward functions, Reinforcement Learning (RL) provides an effective solution to autonomously explore and gradually learn optimal or near-optimal control strategies by iteratively interacting with the environment in the absence of task-specific prior knowledge~\cite{sutton2018reinforcement,kober2013reinforcement}. 
Utilizing the power of deep neural networks~\cite{lecun2015deep} to adapt abstract features from high-dimensional input states, RL has demonstrated superior performances than humans in various complex scenarios including board games~\cite{silver2018general}, video games~\cite{mnih2015human}, and robot control~\cite{ibarz2021train}.
Based on the successful implementations of single-agent RL approaches, people naturally attempt to develop Multi-agent Reinforcement Learning (MARL) to effectively explore optimal control policies of large-scale systems and achieve promising results in a wide range of tasks~\cite{9466421,9119863,oroojlooy2023review,wong2023deep,9466372,9525046,9717997,9765650}.
On the other hand, transferring RL from single-agent environments to multi-agent environments raises a new challenge: the environments affected by the joint actions from multiple agents become non-stationary, and each agent faces a moving-target problem while its optimal strategy strongly depends on the frequently changing policies of other agents.
This characteristic not only breaks the Markov property of the environment but also greatly compromises the learning capability and converge velocity of traditional RL approaches designed for single agent~\cite{busoniu2008comprehensive}.

Compared with the approaches like Independent Q-Learning  (IQL)~\cite{tampuu2017multiagent} that directly implemented single-agent RL approaches in multi-agent scenarios to separately explore independent polices~\cite{matignon2012independent,gupta2017cooperative}, MARL methods based on the Centralized Training with Decentralized Execution (CTDE) framework provides an appealing prospect for addressing the issue of non-stationary~\cite{foerster2016learning}. It enables the multiple agents to learn decentralized policies in a centralized end-to-end fashion: all agents are accessible to the global information during the training stage while the decision-making of each agent is independent during the interaction with the environment.
From the perspective of the value function, Value-Decomposition Networks (VDN)~\cite{sunehag2018value} decomposed the value function for multiple agents under a CTDE framework and achieved better cooperation behaviors in a simulation maze environment.
QMIX~\cite{rashid2020monotonic} further proposed a network-based mixture strategy for multiple agents' value function based on VDN and enjoyed astonishing results performances in StarCraft Multi-Agent Challenge (SMAC)~\cite{samvelyan2019starcraft}.
Based on the policy-based RL approaches, Multi-agent Proximal Policy Optimization (MAPPO)~\cite{yu2022surprising} demonstrated better performances than traditional MARL methods in both Multi-Agent Particle Environment (MPE) and SMAC.
Employing an Actor-Critic (AC) structure, Multi-Agent Deep Deterministic Policy Gradient (MADDPG)~\cite{lowe2017multi} combines the strengths of both value function-based and policy-based approaches and has achieved good learning performances in both cooperative and competitive tasks.
Based on this approach, Multi-agent TD3 (MATD3)~\cite{ackermann2019reducing} tackled the issue of overestimated value function with additional critic networks. The minimax algorithm was further introduced by MiniMax Multi-agent Deep Deterministic Policy Gradient (M3DDPG)~\cite{li2019robust} for enhanced learning capability in both cooperative and competitive tasks.
As demonstrated above, CTDE-based MARL approaches have improved the learning capability of agents from the perspective of structure. However, at the algorithmic level, the inherent inconsistency of multiple agents' policy updates and the resulting deterioration of learning performance has not been sufficiently addressed.

By incorporating the relative entropy between the current and previous policies as a regularization term into the value function, Dynamic Policy Programming (DPP)~\cite{azar2012dynamic} effectively constrains excessive policy updates in single-agent environments. Theoretically, DPP significantly reduces the estimated error of the value function with superior error bounds~\cite{kozuno2019theoretical,vieillard2020leverage}.
In engineering applications of single-agent scenarios, DPP has demonstrated superior sample efficiency and robustness in several robot control tasks~\cite{cui2017kernel,tsurumine2019deep}.
As one of the pioneering works applying relative entropy regularized RL to multi-agent scenarios, Factorial Kernel Dynamic Policy Programming (FKDPP)~\cite{8560593} was proposed to control large-scale chemical plants with multiple DPP agents. It outperformed the control strategy designed by human experts in production rate, profit, and safety on a simulated Vinyl Acetate Monomer (VAM) plant~\cite{zhu2020scalable} and has been successfully implemented in a real-world chemical plant for 35 days~\footnote{This implementation of FKDPP was conducted by Yokogawa Electric Corporation and JSR Corporation. For more details, please see:~\url{https://www.yokogawa.com/news/press-releases/2022/2022-03-22/}.}.
Although this work fully indicates the great potential of relative entropy regularized MARL in real-world systems, FKDPP was developed based on neither deep neural networks nor CTDE framework and was limited in discrete action space without supporting the AC structure. These characteristics restrict its application scope in more challenging and flexible control scenarios.

\begin {table}[t]
\caption{Comparison of MACDPP and the related MARL baselines}
\label{table:1}
\begin{center}
\resizebox{1.0\linewidth}{!}{
\begin{tabular}{|c|c|c|c|c|c|c|c|}
\hline
Approach  & \makecell{Relative Entropy}    & \makecell{AC}    & \makecell{CTDE}  & \makecell{Cooperative \& Competitive} \\ \hline
IQL~\cite{tampuu2017multiagent}  &  $\times$  & $\times$  & $\times$  & $\bigcirc$ \\ \hline
QMIX~\cite{rashid2020monotonic} &  $\times$  & $\times$  & $\times$ & $\times$\\ \hline
MAPPO~\cite{yu2022surprising} & $\times$  & $\times$  & $\bigcirc$ & $\times$\\ 
\hline
MADDPG~\cite{lowe2017multi}&  $\times$  & $\bigcirc$  & $\bigcirc$ & $\bigcirc$\\ \hline
M3DDPG~\cite{li2019robust} & $\times$  & $\bigcirc$  & $\bigcirc$  & $\bigcirc$\\ \hline
MATD3~\cite{ackermann2019reducing} & $\times$  & $\bigcirc$  & $\bigcirc$ & $\bigcirc$\\ 
\hline
FKDPP~\cite{8560593,zhu2020scalable} & $\bigcirc$  & $\times$  & $\times$    & $\times$\\ 
\hline
MACDPP (ours) & $\bigcirc$  & $\bigcirc$  & $\bigcirc$   & $\bigcirc$\\ 
\hline
\end{tabular}}
\end{center}
\end{table}

This paper focuses on integrating the relative entropy regularization to the modern MARL under the CTDE framework in order to alleviate the inconsistency of multiple agents' policy updates in various control scenarios.
According to the characteristics compared with MARL baselines in Table~\ref{table:1}, our proposed approach, Multi-Agent Continuous Dynamic Policy Gradient (MACDPP)\footnote{Code available \url{https://github.com/AdrienLin1/MACDPP}}naturally extends the power of FKDPP from kernel-based value function approximation and discrete action space to the CTDE framework with AC structure with superior learning capability and sample-efficiency.
MACDPP reduces the intractable computational burden of FKDPP in the actor network with continuous actions by Monte Carlo sampling and naturally obtains a superior exploration strategy based on the Boltzmann softmax operator.
Evaluated by both cooperation and competition tasks in MPE environment and OpenAI benchmark control task where multiple agents collaborate to control one high-dimensional system, the proposed method successfully demonstrated superiority in both learning capability and sample-efficiency compared with various multi-agent and signal-agent RL baselines. The contributions of this paper can be summarized as:
\begin{enumerate}
\item Our work first attempts to integrate relative entropy regularization into the CTDE framework-based MARL to address the inherent inconsistency of policy updates for multiple agents at an algorithmic level. We propose a novel MARL approach that is compatible with both cooperative and competitive tasks in a multi-agent scenario, as well as single systems collaboratively controlled by multiple agents.
\item As one natural extension of previous works~\cite{8560593,zhu2020scalable} with successful engineering applications to not only the deep neural networks function approximator but also the AC structure with continuous action space, the proposed MACDPP can be seen as a comprehensive upgrade to FKDPP, targeting enhanced learning capability and control flexibility.  
\item The proposed method was evaluated by several benchmarks from MPE to traditional control tasks in terms of the learning capability and sample efficiency compared to both related CTDE framework-based MARL and widely-applied single-agent RL approaches. We further analyzed the impact of relative entropy regularization in the CTDE framework-based MARL on convergence and control behaviors.
\end{enumerate}

The remainder of this paper is organized as follows. Section~\ref{S2} introduces the preliminaries of Markov games, MARL, and CTDE framework. Section~\ref{S3} details the proposed approach MACDPP. The experimental results are presented in Section~\ref{S4}. Finally, Section~\ref{S5} concludes this paper.

\section{Preliminaries}\label{S2}
\subsection{Markov Games}\label{S2-1}
Markov games are widely utilized to model a  multi-agent environment satisfying partially observable Markov processes (POMDP).
It is generally defined by a sextuple $(N, \mathcal{S}, \mathcal{A}, \mathcal{P}, \mathcal{R}, \gamma)$. $N$ represents the number of agents in the target environment, $\mathcal{S}$ defines the general state space. The locally observed state of the $i$-th agent is denoted as  $\bm{s}_{i}$ which is a subset of the global observed state $ \bm{s}\in\mathcal{S}$.
The joint action is made up of the local actions from all agents $\bm{a}=\bm{a}_{1} \times \cdots \times \bm{a}_{N} \in\mathcal{A}$. The subspace of each agent's action is presented as $\mathcal{A}_i$.
The state transition probability over all agents is presented as $\mathcal{P}$.
$\mathcal{R}$ is a set of reward functions for specific tasks, and each agent has its own reward function $R_{i}(\bm{s}_i, \bm{a}_i, \bm{s}_i^{\prime})\in\mathcal{R}$ based on its local state, action and the next step state. 
The discount factor $\gamma \in [0,1)$ is utilized to gradually ignore the accumulative rewards in a long-term horizon.

Based on the Markov games, MARL introduces the value function $V_{i, \pi_i}$ and the state-action value function $Q_{i, \pi_i}$ to measure the long-term accumulative rewards obtained by the $i$-th agent under its policy:
\begin{gather} 
\begin{split}   
V_{i, \pi_i}(\bm{s}_i) \! = \! \mathbb{E}_{\substack{\bm{s}_{t+1} \sim \mathcal{P},\\ \bm{a}_{i, t} \sim \pi_i}}\!\!\left[\sum_{t=0}^{\infty} \gamma^{t} R_i(\bm{s}_{i, t}, \bm{a}_{i, t}, \bm{s}_{i, t+1}) \mid \bm{s}_{i,0}=\bm{s}_i\right], 
\label{eq_1}
\end{split}
\end{gather}
\begin{gather} 
\begin{split}  
Q_{i, \pi_i}(\bm{s}_i, \bm{a}_i) \! = \! \mathbb{E}_{\substack{\bm{s}_{t+1} \sim \mathcal{P},\\ \bm{a}_{i, t} \sim \pi_i}}\!\!\left[\sum_{t=0}^{\infty} \bm{\gamma}^{t} R_i(\bm{s}_{i, t}, \bm{a}_{i, t}, \bm{s}_{i, t+1}) \mid \substack{\bm{s}_{i,0}=\bm{s}_i,\\ \bm{a}_{i, 0}=\bm{a}_{i}}\right] \label{eq_2}
\end{split}
\end{gather}
where the global state in the next time step $\bm{s}_{t+1}$ is determined by the current global state $\bm{s}_{t}$ and action $\bm{a}_{t}$ under $\mathcal{P}$.
Define $\mathcal{P}_{\bm{s}\bm{s}'}^{\bm{a}}$ as the probability of translating from state $\bm{s}$ to state $\bm{s}^{\prime}$ under action $\bm{a}$ in a global perspective, the goal of each agent in MARL is to learn an optimal control policy to maximize its optimal value function following a Bellman equation:
\begin{align}
V^*_i(\bm{s}_i) = \max_{{\pi}_i} \sum_{\bm{a} \in \mathcal{A}\atop\bm{s}' \in \mathcal{S}} \pi_i(\bm{a}_i|\bm{s}_i)\mathcal{P}_{\bm{s}\bm{s}'}^{\bm{a}}\left(R_i(\bm{s}_i, \bm{a}_i, \bm{s}_i^{\prime}) + \gamma {V^*_i}(\bm{s}^{\prime}_{i})\right).
\label{eq_add_1}
\end{align}

\subsection{Multi-agent Reinforcement Learning in CTDE framework}\label{S2-2}

In the CTDE-framework based MARL like MADDPG~\cite{lowe2017multi}, the AC structure is implemented to separately estimate the state-action value function and model control policy by critic network $\hat{Q}_i(\cdot, \bm{\theta}_i)$ and actor network $\hat{\pi}_i(\cdot, \bm{\phi}_i)$ for each agent where $\bm{\theta}_i$ and $\bm{\phi}_i$ are the corresponding parameters.
In the centralized training process, All critic networks are globally updated with the shared observation information.
Define one global training sample from sample set $\mathcal{D}$ as $(\bm{s}, \bm{a}, \bm{s}^{\prime}, \bm{r})$ where $\bm{r}=[R_1(\bm{s}_1, \bm{a}_1, \bm{s}_1^{\prime}), ..., R_N(\bm{s}_N, \bm{a}_N, \bm{s}_N^{\prime})]$ is the vector of reward signal for all agents, the $i$-th agent's critic networks receive the states and actions from all agents $\hat{Q}_i(\bm{s}, \bm{a}, \bm{\theta}_i)$ and measure its own long-term reward of $R_{i}(\bm{s}_i, \bm{a}_i)$.
Determining the global action of the next step by all agents according to their local states $\bm{a}_i^{\prime} = \hat{\pi}_i(\bm{s}_i^{\prime}, \bm{\phi}_i), \forall i = 1, ..., N$, the corresponding Temporal-Difference (TD) errors that guide the update of critic networks in gradient descent optimization is calculated following:
\begin{gather}
\begin{split}
\mathcal{L}(\bm{\theta}_i)=\left(\hat{Q}_i(\bm{s}, \bm{a}, \bm{\theta}_i)-R_{i}(\bm{s}_i, \bm{a}_i, \bm{s}_i^{\prime}) - \gamma \hat{Q}_i(\bm{s}^{\prime}, \bm{a}^{\prime}, \bm{\theta}_i) \right)^2.
\label{eq_8}
\end{split}
\end{gather}

The actor networks are updated to maximize the returns of the current critic networks based on the local observation of each agent. The corresponding gradient of error is defined as:
\begin{gather}
\begin{split}
\nabla_{\bm{\phi_i}}\leftarrow\nabla_{\bm{a}^{*}_i = \hat{\pi}_i(\bm{s}_i, \bm{\phi}_i)}\hat{Q}_i(\bm{s}, \bm{a}^{*}, \bm{\theta}_i) \nabla_{\bm{\phi}_i}\hat{\pi}_i(\bm{s}_i, \bm{\phi}_i)
\label{eq_10}
\end{split}
\end{gather}
where $\bm{a}^{*}$ is selected by all agents with shared information.
In the decentralized execution process, on the other hand, the control actions of each agent are determined by the actor only without the consideration of other agents.

\section{Approach}\label{S3}
In this section, the proposed method MACDPP was detailed. It naturally extended the relative entropy regularization term from DPP~\cite{azar2012dynamic} and FKDPP~\cite{8560593} to the modern MARL under the CTDE framework and AC structure. 
The multi-agent critic networks regularized by relative entropy were introduced in Section~\ref{S3-2}, and the corresponding actor that supports continuous actions was introduced in Section~\ref{S3-3}. The factorization strategy of MACDPP for cooperative and competitive tasks was discussed in Section~\ref{S3-4} with a summary of MACDPP's learning procedure.

\subsection{Relative Entropy Regularized Critics}\label{S3-2}
Following the existing relative entropy regularized RL approaches~\cite{azar2012dynamic,kozuno2019theoretical}, the difference between the current policy $\pi_i$ and previous policy $\bar{\pi}_i$ of the $i$-th agent on state $\bm{s}_i$ was defined as:
\begin{align}
\mathbb{D}_{\mathrm{KL}}(\bm{s}_i)=\sum_{\bm{a}\in\mathcal{A}} \pi_i(\bm{a}_{i}|\bm{s}_i) \log \left(\frac{\pi_i(\bm{a}_{i}|\bm{s}_i)}{\bar{\pi}_i(\bm{a}_{i}|\bm{s}_i)}\right).
\label{eq_11}
\end{align}
This term was then incorporated into the value function as a regularization term controlled by a parameter $\eta$:
\begin{align}
V_{i}^{\prime}(\bm{s}_i)=\mathbb{E}\left[\sum_{t=0}^{\infty} \gamma^{t}\left(R_i(\bm{s}_{i, t}, \bm{a}_{i, t}, \bm{s}_{i, t+1})-\frac{1}{\eta} \mathbb{D}_{\mathrm{KL}}(\bm{s}_{i, t})\right)\right].
\label{eq_12}
\end{align}
Combining Eqs.\eqref{eq_add_1} and \eqref{eq_12}, the resulted optimal value function was still a Bellman equation with an additional term $-\frac{1}{\eta} \mathbb{D}_{\mathrm{KL}}(\bm{s}_{i, t})$ in  Eq.~\eqref{eq_add_1}.
Assume the action of each agent $\bm{a}_i\in\mathcal{A}_i$ is discrete, an iterative update form of both value function and policy can be found based on DPP~\cite{azar2012dynamic}:
\begin{align}
V_{i, t+1}^{\prime}(\bm{s}_i)  = \frac{1}{\eta} \log \sum_{\bm{a}_i\in\mathcal{A}_i} \exp \left(\eta \cdot \Psi_{i, t}\left(\bm{s}_i, \bm{a}_i\right)\right),
\label{eq_14}
\end{align}
\begin{align}
\bar{\pi}_{i, t+1}(\bm{a}_i|\bm{s}_i) =\frac{\exp\left(\eta\cdot\Psi_{i, t}\left(\bm{s}_i, \bm{a}_i\right)\right)}{\sum_{\bm{a}_i^{\prime}\in\mathcal{A}_i} \exp \left(\eta\cdot \Psi_{i, t}\left(\bm{s}_i, \bm{a}_i^{\prime}\right)\right)}
\label{eq_15}
\end{align}
where $t$ is the iteration of update, $\Psi_{i, t}(\cdot)$ is the action preferences function~\cite{sutton2018reinforcement} which can be treat as a regularized Q function:
\begin{align}
\begin{aligned}
\Psi_{i, t}(\bm{s}_i, \bm{a}_i)  = & \sum_{\bm{s}^{\prime} \in \mathcal{S}} \mathcal{P}_{\bm{s}\bm{s}^{\prime}}^{\bm{a}} \left(R_i(\bm{s}_i, \bm{a}_i, \bm{s}_i^{\prime})+\gamma V_{i, t}^{\prime}\left(\bm{s}^{\prime}_i\right)\right)\\&+\frac{1}{\eta} \log \bar{\pi}_{i, t}(\bm{a}_i | \bm{s}_i).
\label{eq_16}
\end{aligned}
\end{align}

With a discrete action space, once the critic network accurately estimates the action preferences function $\Psi_{i}(\cdot)$, both the value function and policy can be directly calculated. In practice, the transition probability matrix $\mathcal{P}$ is usually too large and inaccessible. DPP proposed an update rule of $\Psi_{i}(\cdot)$ based on sampling by inserting Eqs~\eqref{eq_14} and~\eqref{eq_15} into Eq.~\eqref{eq_16}. Given a sample $(\bm{s}, \bm{a}, \bm{s}^{\prime}, \bm{r})$, the action preferences function of the $i$-th agent was updated following:
\begin{align}
\begin{aligned}
\Psi_{i, t+1}(\bm{s}_i, \bm{a}_i) = &\Psi_{i, t}(\bm{s}_i, \bm{a}_i) - \mathcal{B}_{\eta} \Psi_{i, t}(\bm{s}_i) \\&+ R_i(\bm{s}_i, \bm{a}_i, \bm{s}_i^{\prime}) + \mathcal{B}_{\eta} \Psi_{i, t}(\bm{s}_i^{\prime})
\end{aligned},
\label{eq_19}
\end{align}
\begin{align}
\mathcal{B}_{\eta} \Psi_{i, t}(\bm{s}_i)=\sum_{\bm{a}_i \in \mathcal{A}_i} \frac{\exp \left(\eta \cdot\Psi_{i, t}(\bm{s}_i, \bm{a}_i)\right) \Psi_{i, t}(\bm{s}_i, \bm{a}_i)}{\sum_{\bm{a}^{\prime}_i \in \mathcal{A}_i} \exp \left(\eta \cdot\Psi_{i, t}\left(\bm{s}_i, \bm{a}^{\prime}_i\right)\right)}.
\label{eq_20}
\end{align}
$\mathcal{B}_{\eta}(\cdot)$ is a Boltzmann softmax operator.

It is straightforward to estimate the action preferences function instead of the Q function by the critic neural networks under CTDE framework when the action is discrete. 
Let the critic networks receive the global information of all agents, the loss function of the critic networks $\hat{\Psi}_i(\bm{s}, \bm{a}, \bm{\theta}_i)$ is calculated by integrating Eq.~\eqref{eq_19} into Eq.~\eqref{eq_8}:
\begin{gather}
\begin{split}
\mathcal{L}(\bm{\theta}_i)=\left(\hat{\Psi}_i(\bm{s}, \bm{a}, \bm{\theta}_i)-y(\bm{s}, \bm{a}, \bm{s}_i^{\prime}, \bm{\theta}_i^{-})\right)^2,
\label{eq_21}
\end{split}
\end{gather}
\begin{gather}
\begin{split}
y(\bm{s}, \bm{a}, \bm{s}^{\prime}, \bm{\theta}_i^{-})&=R_{i}(\bm{s}_i, \bm{a}_i, \bm{s}_i^{\prime}) + \hat{\Psi}_i(\bm{s}, \bm{a}, \bm{\theta}_i^{-}) \\&- \mathcal{B}_{\eta}\hat{\Psi}_i(\bm{s}, \bm{\theta}_i^{-})+\gamma\mathcal{B}_{\eta}\hat{\Psi}_i(\bm{s}^{\prime},\bm{\theta}_i^{-})
\label{eq_22}
\end{split}
\end{gather}
Where $\bm{\theta}_i^{-}$ is the parameters of the corresponding target networks. The Boltzmann softmax operator is conducted over the global action of all agents:
\begin{align}
\mathcal{B}_{\eta} \hat{\Psi}_{i}(\bm{s},\bm{\theta}_i^{-})=\sum_{\bm{a} \in \mathcal{A}} \frac{\exp \left(\eta \cdot\hat{\Psi}_{i}(\bm{s}, \bm{a},\bm{\theta}_i^{-})\right) \hat{\Psi}_{i}(\bm{s}, \bm{a},\bm{\theta}_i^{-})}{\sum_{\bm{a}^{\prime} \in \mathcal{A}} \exp \left(\eta \cdot\hat{\Psi}_{i}\left(\bm{s}, \bm{a}^{\prime},\bm{\theta}_i^{-}\right)\right)}.
\label{eq_20_2}
\end{align}

However, the application of the relative entropy regularization in the AC structure remains limited due to the intractable calculation over the whole continuous action space in $\mathcal{B}_{\eta}(\cdot)$. We detailed our solution in the next subsection.

\subsection{Actors with Boltzmann Softmax Sampling}\label{S3-3}

To effectively calculate Eq.~\eqref{eq_20_2} in continuous action space, we estimated it within a local range of the input action in MACDPP. 
The global action $\bm{a}$ was extended to a vector with $M+1$ Monte Carlo samples:
\begin{gather}
\begin{split}
\mathcal{A}^{\text{MC}}=[\bm{a}, \bm{a} + \bm{e}^{1}, \bm{a} + \bm{e}^{2}, ..., \bm{a} + \bm{e}^{M}]
\label{eq_24}
\end{split}
\end{gather}
where $\bm{e}^{j}\sim\mathcal{N}(\bm{0}, \bm{\zeta}^{\text{MC}})$ for $j = 1, ..., M$ is the Monte Carlo sampling noise controlled by $\bm{\zeta}^{\text{MC}}$. The locally estimation of Eq.~\eqref{eq_20} therefore was calculated as:
\begin{align}
\mathcal{B}_{\eta} \hat{\Psi}_{i}(\bm{s}, \bm{\theta}_i^{-}) \approx \sum_{\bm{a} \in \mathcal{A}^{\text{MC}}} \frac{\exp \left(\eta \cdot\hat{\Psi}_{i}(\bm{s}, \bm{a}, \bm{\theta}^{-}_i)\right) \hat{\Psi}_{i}(\bm{s}, \bm{a}, \bm{\theta}^{-}_i)}{\sum_{\bm{a}^{\prime} \in \mathcal{A}^{\text{MC}}} \exp \left(\eta \cdot\hat{\Psi}_{i}\left(\bm{s}, \bm{a}^{\prime}, \bm{\theta}_{i}^{-}\right)\right)}
\label{eq_25}
\end{align}
According to~\cite{asadi2017alternative}, one critical issue of the Boltzmann softmax operator in RL is its multiple fixed points without non-expansion property which guarantees the convergence of ”Q-learning like” algorithms to a unique fixed point in theory. One effective solution is to replace it with the Mellowmax operator with a unique fix point and non-expansion property:
\begin{align}
\mathcal{M}_{\eta} \hat{\Psi}_{i}(\bm{s}, \bm{\theta}_i^{-}) \approx \frac{1}{\eta}\log\left(\frac{\sum_{\bm{a} \in \mathcal{A}^{\text{MC}}}\exp\left(\eta \cdot\hat{\Psi}_{i}(\bm{s}, \bm{a}, \bm{\theta}^{-}_i)\right)}{M+1}\right).
\label{eq_26}
\end{align}
Algorithm~\ref{a1} summarized the calculation of $\mathcal{M}_{\eta} \hat{\Psi}_{i}(\bm{s}, \bm{\theta}_i^{-})$.
In practice, numerical issues usually arise in Eq.~\eqref{eq_26} with a large $\eta$. We alternatively calculated it following:
\begin{align}
\begin{split}
& \mathcal{M}_{\eta} \hat{\Psi}_{i}(\bm{s}, \bm{\theta}_i^{-}) \approx \\&\frac{1}{\eta}\log\left(\frac{\sum_{\bm{a} \in \mathcal{A}^{\text{MC}}}\exp\left(\eta \cdot\hat{\Psi}_{i}(\bm{s}, \bm{a}, \bm{\theta}^{-}_i)-C\right)}{M+1}\right)+C
\label{eq_26_2}
\end{split}
\end{align}
where $C=\eta\cdot\max_{\bm{a}^{\prime} \in \mathcal{A}^{\text{MC}}}[\hat{\Psi}_{i}(\bm{s}, \bm{a}^{\prime}, \bm{\theta}^{-}_i)]$.

Employing the Monte Carlo sampling to estimate $\mathcal{M}_{\eta} \hat{\Psi}_{i}(\cdot)$, any policy network maps the local states to the local actions can be used as an actor.
In the decentralized execution process, the trained agent made decisions based on its own actor with local observation $\bm{a}_i^{*}=\hat{\pi}_i(\bm{s}_i, \bm{\phi}_i)$
The actor was updated in the centralized training following the gradient below:
\begin{gather}
\begin{split}
\nabla_{\bm{\phi_i}}\leftarrow\nabla_{\bm{a}^{*}_i = \hat{\pi}_i(\bm{s}_i, \bm{\phi}_i)}\hat{\Psi}_i(\bm{s}, \bm{a}^{*}, \bm{\theta}_i) \nabla_{\bm{\phi}_i}\hat{\pi}_i(\bm{s}_i, \bm{\phi}_i)
\label{eq_10_2}
\end{split}
\end{gather}
where the $\bm{a}^{*}$ was jointly calculated by all agents' policies.

\begin{algorithm}[t]
	\caption{Estimation by Monte Carlo Sampling}
	\SetKwFunction{CM}{MC\_Estimate}
	\SetKwProg{Fn}{Function}{:}{}
    \Fn{\CM{$\bm{s}, \bm{a}, \bm{\theta}_i$}}{
    \# Build Monte Carlo sample set\\
	\For{$m=1$ to $M$}{$\bm{e}^{m}\sim\mathcal{N}(\bm{0}, \bm{\zeta})$}
	$\mathcal{A}^{\text{MC}}=[\bm{a}, \bm{a} + \bm{e}^{1}, \bm{a} + \bm{e}^{2}, ..., \bm{a} + \bm{e}^{M}]$\\
    \# Locally estimate Mellowmax operation\\
    $\mathcal{M}_{\eta} \hat{\Psi}_{i}(\bm{s}, \bm{\theta}_i^{-}) \approx \frac{1}{\eta}\log\left(\frac{\sum_{\bm{a} \in \mathcal{A}^{\text{MC}}}\exp\left(\eta \cdot\hat{\Psi}_{i}(\bm{s}, \bm{a}, \bm{\theta}^{-}_i)-C\right)}{M+1}\right)+C$\\
    Return $\mathcal{M}_{\eta}\hat{\Psi}_{\text{target}}(\bm{s}, \bm{\theta}^{-})$
}
\label{a1}
\end{algorithm}

\begin{algorithm}[t]
	\caption{Exploration using Boltzmann Softmax}
	\SetKwFunction{CB}{BS\_Sampling}
	\SetKwProg{Fn}{Function}{:}{}
    \Fn{\CB{$\bm{s}, \bm{\theta}_i$}}{
    \For{$j=1$ to $N$}{
    $\bm{a}_{j}^{*}=\hat{\pi}_{j}(\bm{s}, \bm{\phi}_{j})$\\
    }
	$\bm{a}^{*} = [\bm{a}_{1}^{*}, ..., \bm{a}_{N}^{*}]$\\
	\For{$m=1$ to $M$}{
	$\bm{e}^{m}_i\sim\mathcal{N}(\bm{0}, \bm{\zeta}^{\text{Sampling}}), \bm{e}^{m}_{j\neq i} = \bm{0}$\\
	$\bm{e}^{m}=[\bm{e}^{m}_1, ..., \bm{e}^{m}_N]$\\
	}
	$\mathcal{A}^{\text{explore}}=[\bm{a}^{*}, \bm{a}^{*} + \bm{e}^{1}, \bm{a}^{*} + \bm{e}^{2}, ..., \bm{a}^{*} + \bm{e}^{M}]$\\
	Select $\bm{a}^{\text{explore}}\in\mathcal{A}^{\text{explore}}$ following:\\
	$p(\bm{a}^{\text{explore}}) = \frac{\exp \left(\eta \cdot\hat{\Psi}_{i}(\bm{s}, \bm{a}^{\text{explore}}, \bm{\theta}^{-}_i)\right)}{\sum_{\bm{a}^{\prime} \in \mathcal{A}^{\text{explore}}} \exp \left(\eta \cdot\hat{\Psi}_{i}\left(\bm{s}, \bm{a}^{\prime}, \bm{\theta}_{i}^{-}\right)\right)}$\\
	\Return{$\bm{a}^{\text{explore}}_i$}
}
\label{a2}
\end{algorithm}

Unlike MADDPG which explores by directly adding noises to its actions, MACDPP proposed an effective exploitation that naturally related to the relative entropy regularized value function based on the shared information in the centralized training process.
Define the global action as $\bm{a}^{*}=[\bm{a}_1^{*}, ..., \bm{a}_N^{*}]$ where 
$\bm{a}_i^{*}=\hat{\pi}_i(\bm{s}_i, \bm{\phi}_i), i = 1, ...N$.
An exploration set for the $i$-th agent with $M+1$ candidates was built following Algorithm~\ref{a2}: $\mathcal{A}^{\text{explore}}=[\bm{a}^{*}, \bm{a}^{*} + \bm{e}^{1}, \bm{a}^{*} + \bm{e}^{2}, ..., \bm{a}^{*} + \bm{e}^{M}]$ where exploration noises $\bm{e}^{m}$ added Gaussian noises only to the local actions related to the $i$-th agent. Please note that the variance of sampling $\bm{\zeta}^{\text{Sampling}}$ which affected the decision-making of the agent was independent of $\bm{\zeta}^{\text{MC}}$ in Algorithm.~\ref{a1} which locally estimated the Mellowmax operation.
An effective exploration action was randomly selected following the probability below:
\begin{align}
p(\bm{a}^{\text{explore}}) = \frac{\exp \left(\eta \cdot\hat{\Psi}_{i}(\bm{s}, \bm{a}^{\text{explore}}, \bm{\theta}^{-}_i)\right)}{\sum_{\bm{a}^{\prime} \in \mathcal{A}^{\text{explore}}} \exp \left(\eta \cdot\hat{\Psi}_{i}\left(\bm{s}, \bm{a}^{\prime}, \bm{\theta}_{i}^{-}\right)\right)}.
\label{eq_27}
\end{align}
The $i$-th agent utilized the corresponding local action $\bm{a}^{\text{explore}}_i$ in $\bm{a}^{\text{explore}}$ to interact with the environment.
Although Eq.~\eqref{eq_27} required the global actions of all agents for an effective exploration, the execution of all agents can be decentralized in evaluation since the deterministic local action $\bm{a}_i^{*}=\hat{\pi}_i(\bm{s}_i, \bm{\phi}_i)$ was related to only the local observation $\bm{s}_i$.

\begin{algorithm}[t]
\caption{Learning Process of MACDPP}
\For{$i=1$ to $N$}{
Initialize buffer $D_i$, networks weights $\bm{\theta}_i$, $\bm{\phi}_i$.\\
Copy the target networks with parameters $\bm{\theta}_i^-$, $\bm{\phi}_i^-$.\\
}
\For{$e=1$ to $E$}{
    \For{$t=1$ to $T$}{
		\# Interaction phase \\
		Observe state $\bm{s}$\\
        \For{$i=1$ to $N$}{
		$\bm{a}_i = \text{BS\_Sampling}(\bm{s}, \bm{\theta}^{-}_i)$}
		Execute $\bm{a}=[\bm{a}_{1}, ..., \bm{a}_{N}]$\\
		Observe next state $\bm{s}^{\prime}$ and reward $\bm{r}$\\
		Separately store sample to $\mathcal{D}_i, i = 1, ..., N$\\
		\# Centralized training phase \\
        \For{$i = 1$ to $N$}{
            Sample mini-batch of $J$ samples from $\mathcal{D}_{i}$\\
            \For{$j = 1$ to $J$}{
			\# Restore information from $\mathcal{D}_{k}, k\neq i$\\
			$\bm{s}_j=[\bm{s}_{1,j}, ..., \bm{s}_{N,j}]$\\
			$\bm{a}_j=[\bm{a}_{1,j}, ..., \bm{a}_{N,j}]$\\
			$\bm{s}_j^{\prime}=[\bm{s}_{1,j}^{\prime}, ..., \bm{s}_{N,j}^{\prime}]$\\
            \# Calculate the next action of all agents \\
                \For{$k = 1$ to $N$}{
                $\bm{a}_{k,j}^{\prime} = \hat{\pi}_{k}(\bm{s}_{k}^{\prime}, \bm{\phi}^{-}_{k})$\\}
			$\bm{a}^{\prime}_j=[\bm{a}_{1,j}^{\prime}, ..., \bm{a}_{N,j}^{\prime}]$\\
			\# Monte Carlo Estimation \\
            $\mathcal{M}_{\eta} \hat{\Psi}_{i}(\bm{s}_{j}) = \text{MC\_Estimate}(\bm{s}_{j}, \bm{a}_{j},\bm{\theta}_i^{-})$\\
            $\mathcal{M}_{\eta} \hat{\Psi}_{i}(\bm{s}_{j}^{\prime}) = \text{MC\_Estimate}(\bm{s}_{j}^{\prime}, \bm{a}_{j}^{\prime}, \bm{\theta}_i^{-})$\\
            \# Calculate TD error following Eq.~\eqref{eq_22}\\
            $y_{j} = R_i(\bm{s}_{i,j}, \bm{a}_{i,j}, \bm{s}_{i,j}^{\prime}) + \gamma \mathcal{M}_{\eta} \hat{\Psi}_{i}(\bm{s}_{j}^{\prime}) + \hat{\Psi}_{i}(\bm{s}_{j}, \bm{a}_{j}, \bm{\theta}^{-}_i) - \mathcal{M}_{\eta} \hat{\Psi}_{i}(\bm{s}_{j})$\\
            }

        \# Update critic $\hat{\Psi}_i(\cdot, \bm{\theta}_{i})$ and actor $\hat{\pi}_i(\cdot, \bm{\phi}_{i})$\\
        $\bm{\theta}_{i} \leftarrow \operatorname*{arg\,min}\limits_{\bm{\theta}_{i}}\frac{1}{J}\sum_{j=1}^{J}(y_{j} - \hat{\Psi}(\bm{s}_j,\bm{a}_{i,j}, \bm{\theta}_{i}))^2$\\
        \# Update actor $\hat{\pi}_i(\cdot, \bm{\phi}_{i})$\\
        $\nabla_{\bm{\phi}_{i}} \!\!\leftarrow\!\!\frac{\sum_{j=1}^{J} \!\!\!\nabla_{\bm{a}^{\prime}_{i,j}} \!\!\!\!\hat{\Psi}(\bm{s}_{j}, \bm{a}^{\prime}_{j}, \bm{\theta})\!\nabla_{\bm{\phi}_i} \!\!\hat{\pi}_i(\bm{s}_{i, j}, \bm{\phi}_i)}{J}$\\
		\# Update target networks\\
		$\bm{\theta}^{-}_{i} \leftarrow \tau \bm{\theta}_{i} + (1-\tau)\bm{\theta}^{-}_{i}$\\
		$\bm{\phi}^{-}_{i} \leftarrow \tau\bm{\phi}_{i} + (1-\tau)\bm{\phi}^{-}_{i}$\\
        }
}
}
\Return{$\hat{\Psi}_i, \hat{\pi}_i, i = 1, ..., N$}
	\label{a3}
\end{algorithm}

\subsection{Factorization of Multi-agents in Different Tasks}\label{S3-4}

In this subsection, we detailed the training procedure of MACDPP in both multi-agent cooperative/competitive environments and single systems that are collaboratively controlled by multiple agents.
The learning process of the proposed method in a multi-agent cooperative/competitive scenario was summarized in Algorithm~\ref{a3}.
Given the length of episode $E$ and the length of one rollout $T$, 
at the beginning, the parameters of both critic and actor networks were randomly initialized as $\bm{\theta}_i, \bm{\phi}_i, i = 1, ..., N$.
Those weights were copied to the target networks as $\bm{\theta}_i^{-}, \bm{\phi}_i^{-}$.
At each step, the global state $\bm{s}$ was first observed. The control action of each agent $\bm{a}_i$ was determined by $\text{BS\_Sampling}(\cdot)$ following Algorithm~\ref{a2}.
Conducted global action $\bm{a}=[\bm{a}_{1}, ..., \bm{a}_{N}]$ by all agents, the global state in the next step and the vector of $N$ reward functions $\bm{r}=[R_1(\bm{s}_1, \bm{a}_1, \bm{s}_1^{\prime}), ..., R_N(\bm{s}_N, \bm{a}_N, \bm{s}_N^{\prime})]$ were then observed and stored to the separate replay buffer.
During the centralized training buffer, the update of each agent was separately conducted with its own $J$ mini-batch samples while the samples from other agents' buffers were used to restore the global information.
The TD error was calculated following Eq.~\eqref{eq_21} and Algorithm~\ref{a1} to updated actor and critic networks. 
The target networks were then smoothly updated with a smooth parameter $\tau$ according to $\bm{\theta}_i, \bm{\phi}_i, i = 1, ..., N$.

\begin{figure}[t]
\centering
\includegraphics[width=1.0\columnwidth]{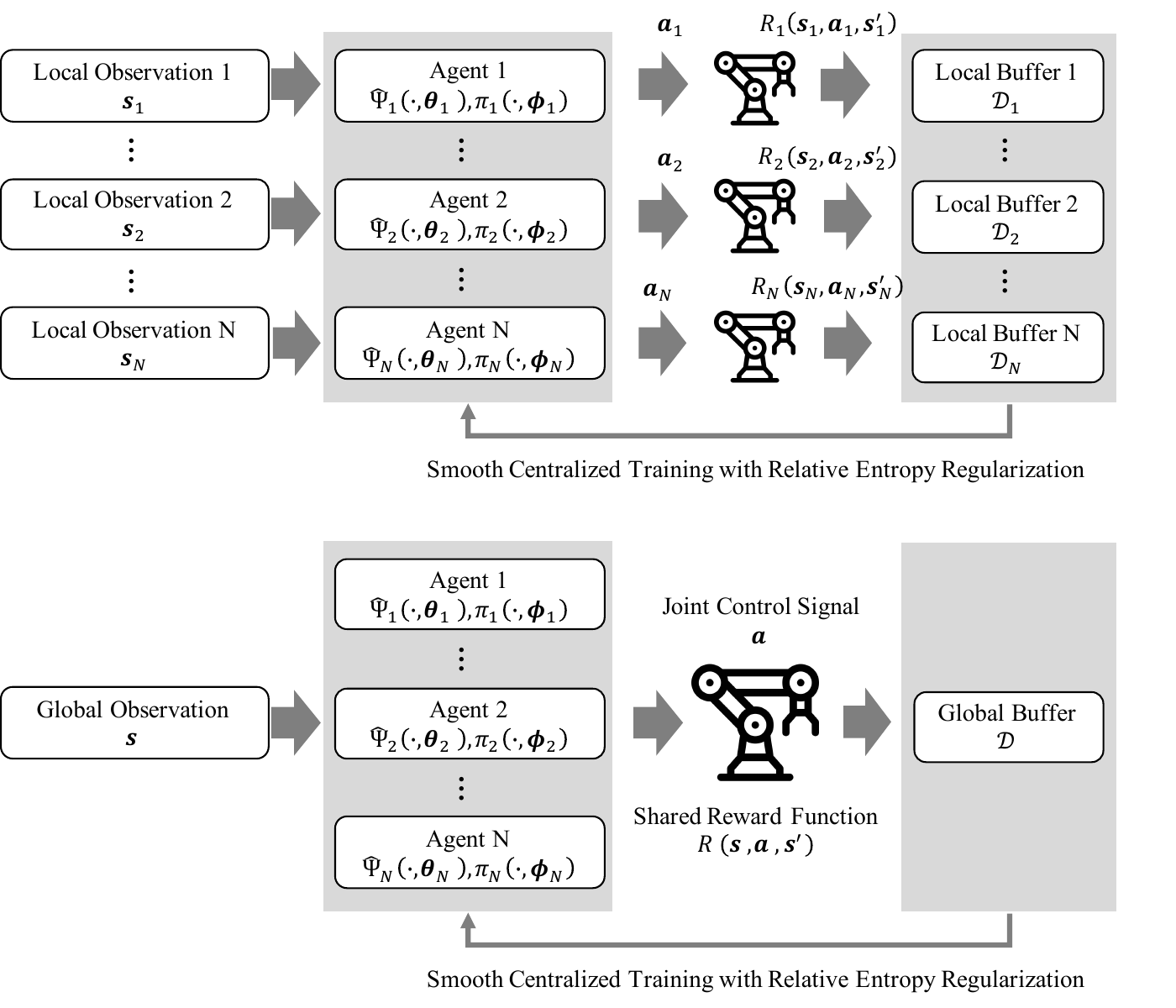}
\caption{The differences between the learning process of MACDPP in cooperative/competitive (top) and joint control (bottom) scenarios.}
\label{figure:diff}
\end{figure}

When implementing the proposed MACDPP to jointly control one complex system by multiple agents following our previous work~\cite{8560593,zhu2020scalable}, only one global replay buffer $\mathcal{D}$ was built.
At each step, the global observed state $\bm{s}$ was sent in parallel to all agents. The control actions of all actors were then integrated as $\bm{a}$ and conducted to the target system. The resulting next step state $\bm{s}^{\prime}$ and the corresponding reward were received and stored in $\mathcal{D}$.
Please note that all agents shared one reward function $\bm{r}=[R(\bm{s}, \bm{a}, \bm{s}^{\prime})]$ designed for the whole system.
Unlike the case in charge of multi-agent cooperative/competitive environments, MACDPP did not separately conduct $J$ mini-batch sampling for each agent when jointly controlling one system but rather shared $J$ samples during the update.
In addition, the actor network of the $i$-th agent $\hat{\pi}_i(\bm{s}, \bm{\phi}_i)$ directly received the global observed state in the decision-making process. The difference between the learning process of MACDPP in cooperative/competitive and joint control scenarios was illustrated in Fig.~\ref{figure:diff}.

\begin{figure}[t]
\centering
\includegraphics[width=1.0\columnwidth]{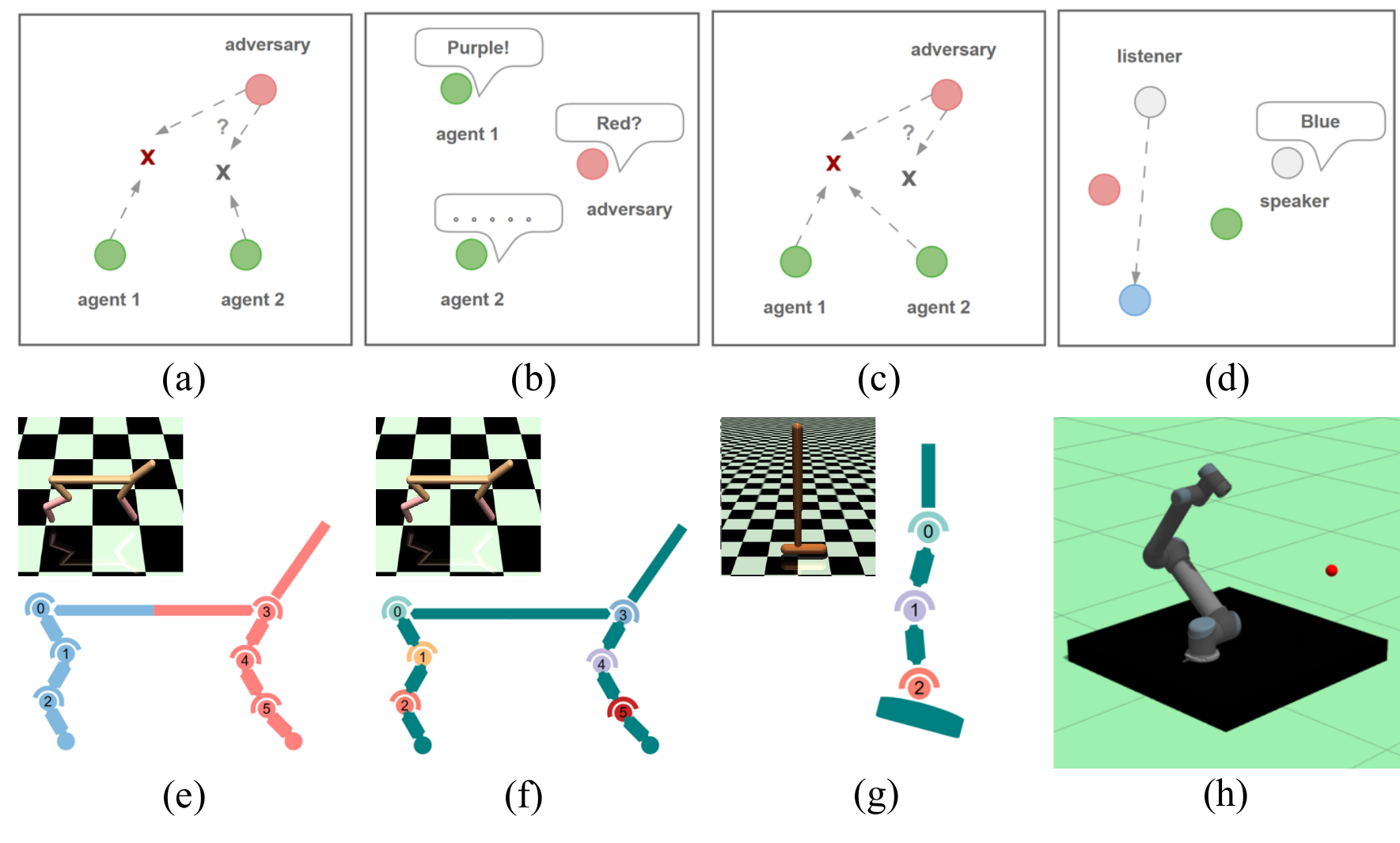}
\caption{Eight Benchmark tasks for evaluation: (a) Physical Deception from MPE (mixed cooperative competitive); (b) Covert Communication from MPE (mixed cooperative competitive); (c) Keep Away from MPE (mixed cooperative competitive); (d) Cooperative Communication from MPE (pure cooperative); (e) HalfCheetah from Mujoco (joint control by two agents); (f) HalfCheetah from Mujoco (joint control by six agents); (g) Hopper from Mujoco (joint control by three agents); (h) UR5 End Effector Positioning from robo-gym (joint control by five agents)}
\label{figure:benchmark}
\end{figure}

\section{Experimental Results}\label{S4}

\subsection{Experimental Settings}\label{S4-1}

\begin{table*}[t]
	\caption{Common hyperparameter settings of compared approaches}
	\centering
	\label{table:2}
	\resizebox{1.0\linewidth}{!}{\begin{tabular}{|l|c|c|c|c|c|c|c|}
        \hline
		&\text{Physical Deception}&\text{Covert Communication}&\text{Keep Away}&\text{Cooperative Communication}&\text{HalfCheetah}&\text{Hopper}&\text{UR5}\\\hline
		Critic Learning Rate &$0.01$ & $0.01$ & $0.1$ & $0.1$ & $10^{-3}$ & $5\times10^{-4}$ & $5\times10^{-4}$\\\hline
		  Actor Learning Rate &$0.01$ & $0.01$ & $0.01$ & $0.01$ & $10^{-3}$ & $5\times10^{-5}$ & $5\times10^{-5}$ \\\hline
	      Actor and Critic Structure &$(64,64)$ &$(64,64)$ & $(64,64)$ & $(64,64)$  &$(400,300)$ & $(400,300)$ & $(400,300)$ \\\hline
	      Target Update Rate ($\tau$) & $1\times10^{-3}$ & $1\times10^{-4}$ & $1\times10^{-4}$  & $1\times10^{-4}$ & $5\times10^{-3}$ & $5\times10^{-3}$ & $5\times10^{-3}$\\\hline
            Batch Size ($J$) &$1024$ &$1024$ &$1024$ &$1024$ &$100$ & $100$ & $100$\\\hline
            Discount Factor ($\gamma$) &0.95 &0.95 &0.95 &0.95 &0.99 &0.99  &0.99\\\hline
            Memory Size &$10^{5}$ &$10^{5}$ &$10^{5}$ &$10^{5}$ &$10^{6}$ & $10^{6}$ & $10^{6}$\\\hline
            Warmup Steps & $0$ &$0$ &$0$ &$ 0 $ &$10^{4}$ &$10^{4}$ &$10^{4}$\\\hline
			Steps per Update & $100$ &$50$ &$25$ &$25$ &$1$ &$1$ &$1$\\\hline
	\end{tabular}}
\end{table*}

\begin{table}[t]
	\caption{Specific Hyperparameters of MACDPP}
	\centering
	\label{table:3}
	\resizebox{1.0\linewidth}{!}{\begin{tabular}{|c|c|c|c|c|c|}
        \hline
		&$\eta$&$M_1$ &$M_2$&$\bm{\zeta}^{\text{MC}}$& $\bm{\zeta}^{\text{Sampling}}$\\\hline
	Physical Deception & $20$ &$30$ &$50$&$\bm{0.1}$&$\bm{0.2}$\\\hline
	Covert Communication &$20$   &$30$& $50$&$\bm{0.1}$&$\bm{0.2}$\\\hline
    Keep Away &$0.1$  &$30$& $50$&$\bm{0.1}$&$\bm{0.1}$  \\\hline
    Cooperative Communication  &$0.1$  &$30$& $50$&$\bm{0.1}$&$\bm{0.1}$\\\hline
	HalfCheetah &$20$  &$30$& $50$&$\bm{0.1}$&$\bm{0.1}$\\\hline
	Hopper &$5$  &$30$& $50$&$\bm{0.1}$&$\bm{0.1}$\\\hline
	UR5 &$0.05$  &$30$& $50$&$\bm{0.1}$&$\bm{0.1}$\\\hline
	\end{tabular}}
\end{table}

In this section, we evaluated MACDPP in multi-agent and traditional control tasks in terms of learning capability and sample efficiency.
For the multi-agent scenario, we selected the physical deception, Covert Communication, keep away and cooperative communication tasks from the Multi-Agent Particle Environment (MPE)~\cite{lowe2017multi}\footnote{\url{https://github.com/openai/multiagent-particle-envs}.} The first three are mixed cooperative-competitive tasks and the last one is a pure cooperative task.
MADDPG~\cite{lowe2017multi}, MATD3~\cite{ackermann2019reducing} and M3DDPG~\cite{li2019robust} were selected as the compared MARL baselines.
For the traditional control scenario, we selected the Ant, HalfCheetah from Mujoco simulation~\cite{todorov2012mujoco} and the UR5 robot arm simulation task ur\_ee\_position developed in robo-gym~\cite{lucchi2020robo}. For each MARL approach, the HalfCheetah was two scenarios: 2 agents separately controlled the front and back body, and six agents controlled six joints. The Hopper and UR5 were jointly controlled by three and five agents for each controllable joint.
We not only compared the proposed method with MARL approaches MADPPG~\cite{lowe2017multi} but also the widely implemented single-agent RL approaches including Deep Deterministic Policy Gradient (DDPG)~\cite{lillicrap2016continuous}, Twin Delayed Deep Deterministic Policy Gradient (TD3)~\cite{fujimoto2018addressing} and Soft Actor-Critic (SAC)~\cite{haarnoja2018soft}.
All benchmark control tasks were illustrated in Fig.~\ref{figure:benchmark}.
The hyperparameters of all compared methods for each task were summarized in Table~\ref{table:2}. All actors and critics shared the same network structures. The tunable hyperparameters of the proposed MACDPP including $\eta$, the Monte Carlo sampling numbers in Algorithms~\ref{a1} and ~\ref{a2} $M_1, M_2$ and the sampling noise $\bm{\zeta}^{\text{MC}}, \bm{\zeta}^{\text{Sampling}}$ were listed in Table~\ref{table:3}.
The proposed MACDPP was developed by PaddlePaddle~\cite{ma2019paddlepaddle} under its RL toolkit PARL~\footnote{\url{https://github.com/PaddlePaddle/PARL}.}.
All experiments were conducted on a workstation with Intel Xeon W2265 CPU, NVIDIA GeForce RTX 3080 GPU, 64GB memory and Ubuntu 20.04 OS. 
The experimental results were summarized over five independent trials with different random seeds for statistical evidence.

\subsection{Cooperation and Competition in MPE Benchmarks}\label{S4-2}
\subsubsection{Evaluation of the Learning Capability}\label{S4-2-1}

\begin{table*}
	\caption{Maximum Average Returns over four MPE benchmark tasks}
	\centering
	\label{table:lr1}
        \resizebox{0.75\linewidth}{!}{
	\begin{tabular}{|l|c|c|c|c|}
        \hline
		&\textbf{MACDPP}&\textbf{MADDPG}&\textbf{MATD3}&\textbf{M3DDPG}\\\hline
		Physical Deception & {\color[HTML]{C73656} $\bm{47.02\pm14.28}$} & $-0.36\pm2.20$   & $0.67\pm2.63$ & $3.77\pm0.23$\\\hline
		Covert Communication & {\color[HTML]{C73656} $\bm{31.68\pm19.07}$} & $0.72\pm2.71$ & $1.76\pm8.78$ & $13.48\pm10.52$\\\hline
		Keep Away & {\color[HTML]{C73656} $\bm{-5.27\pm1.27}$} & $-6.27\pm1.01$    & $-5.55\pm1.33$   & $-8.53\pm0.10$\\\hline
	    Cooperative Communication & $-13.10\pm6.79$  & $-15.88\pm10.78$  &  {\color[HTML]{C73656} $\bm{-11.90\pm6.95}$} & $-40.07\pm11.90$ \\\hline
	\end{tabular}}
\end{table*}

\begin{figure*}[t]
\centering
\includegraphics[width=1.5\columnwidth]{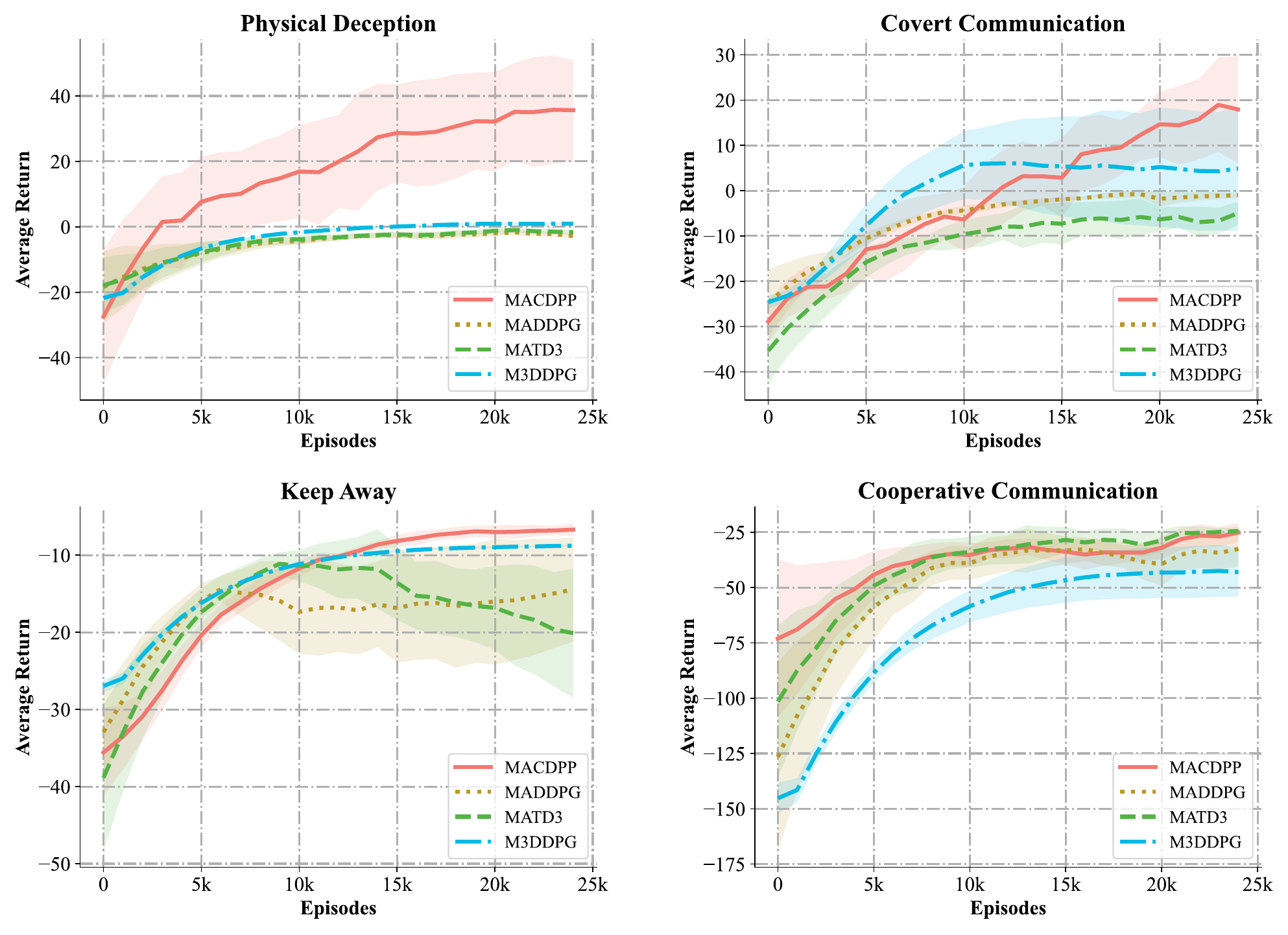}
\caption{Learning curves of MACDPP and other MARL baselines in MPE benchmark tasks. The shaded region represents the corresponding standard deviation over five trials.}
\label{figure:lr1}
\end{figure*}

We first compared the proposed methods with the related MARL approaches in four benchmark tasks from the MPE environment. The learning curves of all compared approaches were shown in Fig.~\ref{figure:lr1} while the maximum average returns of each method in the evaluation phase during the learning were listed in Table~\ref{table:lr1} (the number in red indicates the best result in the corresponding term).
The trials of all four tasks were conducted by $25$k episodes, each episode had $25$ steps.

In the Physical Deception task, MADDPG, MATD3 and M3DDPG converged to close performances near $0$ average return after $25$k episodes.
As a comparison, our method quickly suppressed other baselines in the first $2000$ episodes and converged to $40$ average return after $25$k episodes.
In the evaluation task, MACDPP outperformed MATD3 and M3DDPG with over $7000\%$ and $1200\%$ maximum average return while MADDPG achieved a negative average return.
In the Covert Communication task, M3DDPG outperformed MADDPG and MATD3 in both average return and converge velocity thanks to its Minimax operator.
On the other hand, MACDPP converged to over $100\%$ more average return during learning and achieved $135\%$ maximum average in evaluation compared with the suboptimal method M3DDPG.
In the Keep Away task, MADDPG converged to a relatively low average return. MATD3 converged to near $-10$ average return within $10$k episodes but could not maintain its performance. Although M3DDPG converged quickly in the first $10$k episodes, MACDPP outperformed it with the best maximum average return and the lowest standard deviation in the learning curve which indicated a more stable training procedure.
In the Cooperative Communication task which requires pure cooperation over all agents, it is observed that M3DDPG which is good at competition failed to learn good cooperation policies. Although MACDPP, MADDPG and MATD3 all learned to a close performance, the proposed method enjoyed the fastest convergence velocity.
Overall, the proposed method demonstrated significantly superior learning capability than related MARL baselines in various MPE benchmark tasks.

\subsubsection{Evaluation of the Sample Efficiency}\label{S4-2-2}

\begin{figure}[h]
\centering
\includegraphics[width=0.85\columnwidth]{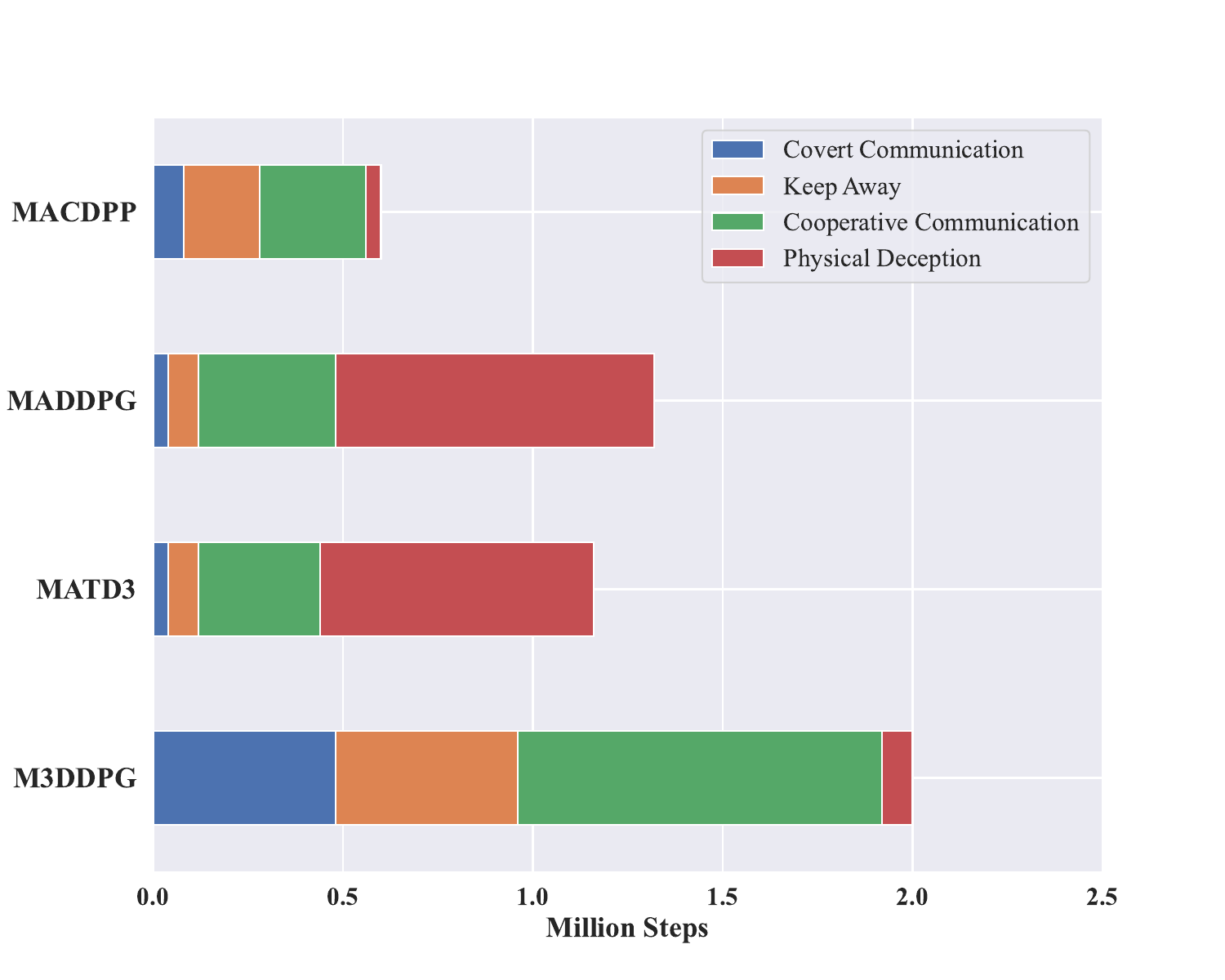}
\caption{Number of interactions utilized by all compared approaches in MPE benchmark tasks to reach the lower boundary of maximum average return.}
\label{figure:se1}
\end{figure}

\begin{table*}[t]
	\caption{Average calculation time of $1000$ episodes over four MPE benchmark tasks}
	\centering
	\label{table:time1}
        \resizebox{0.75\linewidth}{!}{
	\begin{tabular}{|l|c|c|c|c|}
        \hline
		&\textbf{MACDPP}&\textbf{MADDPG}&\textbf{MATD3}&\textbf{M3DDPG}\\\hline
		Physical Deception & $124.77\pm1.77$ s & $32.84\pm0.20$ s   & $33.31\pm0.20$ s & $105.86\pm2.41$ s\\\hline
		Covert Communication & $161.73\pm1.40$ s & $41.23\pm0.35$ s   & $41.27\pm0.45$ s & $125.83\pm2.38$ s\\\hline
		Keep Away & $149.58\pm1.33$ s & $39.30\pm0.32$ s   & $42.94\pm0.39$ s & $115.28\pm1.86$ s\\\hline
	    Cooperative Communication & $144.11\pm0.55$ s & $37.10\pm0.23$ s   & $39.19\pm0.41$ s & $108.56\pm2.44$ s\\\hline
	\end{tabular}}
\end{table*}

\begin{table*}
	\caption{Maximum Average Returns over four benchmark control tasks}
	\centering
	\label{table:lr2}
        \resizebox{1.0\linewidth}{!}{
	\begin{tabular}{|l|c|c|c|c|c|}
        \hline
		&\textbf{MACDPP}&\textbf{MADDPG}&\textbf{DDPG}&\textbf{TD3} &\textbf{SAC}\\\hline
		HalfCheetah (2 agents) & {\color[HTML]{C73656} $\bm{10590.80\pm443.92}$} & $9404.49\pm1252.08$   & $9991.71\pm1533.64$ & $9754.40\pm955.57$ & $10426.10\pm1577.89$\\\hline
		HalfCheetah (6 agents) & $9767.99\pm419.40$ & $6482.54\pm1296.47$ & $9821.45\pm1393.96$ & $9730.61\pm962.49$ & {\color[HTML]{C73656} $\bm{9828.28\pm803.50}$} \\\hline
		Hopper (3 agents) & {\color[HTML]{C73656} $\bm{3452.87\pm60.85}$} & $2741.40\pm935.69$    & $2212.34\pm1139.88$   & $3384.24\pm230.43$ & $3325.33\pm270.24$\\\hline
	    UR5 (5 agents) & {\color[HTML]{C73656} $\bm{1.86\pm0.03}$} & $1.66\pm0.73$   & $0.60\pm1.04$ & $1.60\pm0.30$ & $-1.13\pm0.19$\\\hline
	\end{tabular}}
\end{table*}

The sample efficiency which is important to the implementation of RL in real-world systems was evaluated in Fig.~\ref{figure:se1}. We define the measure of sample efficiency in this subsection as the number of interactions used by each approach to reach the lower boundary of the maximum average returns in Table~\ref{table:lr1} over the four benchmark tasks from the MPE environment.
It is clearly observed that the proposed method achieved the overall superior sample efficiency among all compared MARL approaches, it reduced $70\%$, $54.6\%$ and $48.3\%$ usage of samples than M3DDPG, MADDPG and MATD3 to reach a certain level of control performances.
At the same time, we found that MACDPP has overall effectiveness in improving sample efficiency, whether in cooperative or competitive tasks.
This result demonstrated the great potential of the proposed method in quickly learning proper multi-agent control policies in complex scenarios with few sampling costs.

\subsubsection{Evaluation of the Computational Efficiency}\label{S4-2-3}

In this subsection, we investigated the impact of additional Monte Carlo sampling and Boltzmann softmax operator in MACDPP on computational efficiency. We measured the average calculation time of $1000$ episodes over all compared MARL baselines in Table~\ref{table:time1}. It is observed that MACDPP brought additional computational burdens in all four tasks. It required $285.54\%$, $270.19\%$ and $27.36\%$ more calculation time compared with MADDPG, MATD3 and M3DDPG.
On the other hand, considering the obvious advantages of our approach in learning ability, convergence velocity, and sample efficiency, we believed that these increased computational complexities in training and decision-making were acceptable.

\begin{figure}
\centering
\includegraphics[width=0.8\columnwidth]{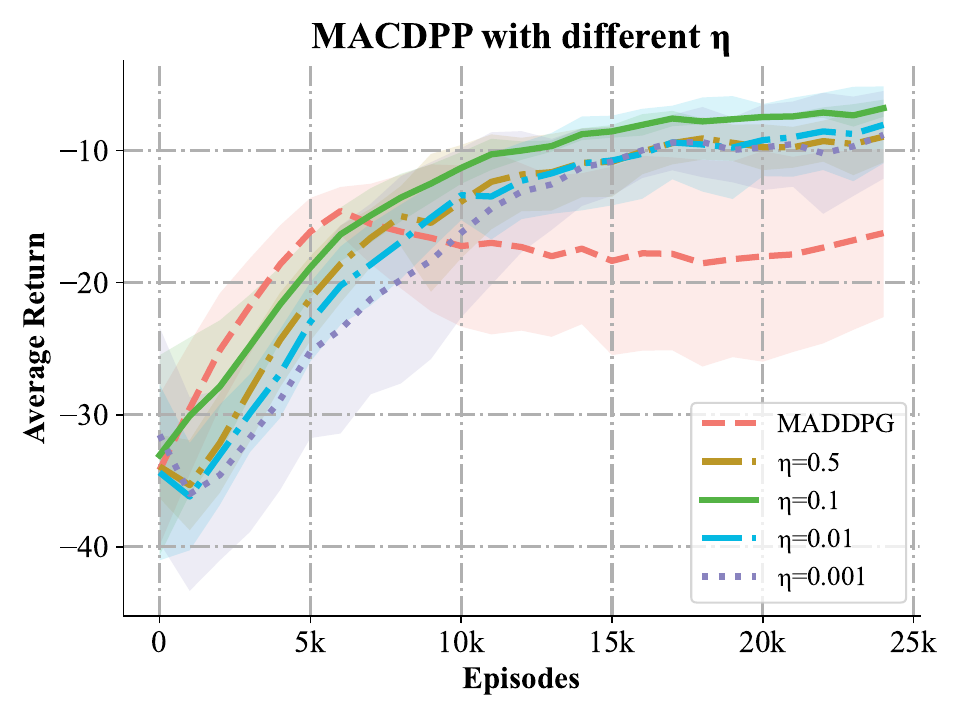}
\caption{Learning curves of MACDPP with different values of parameter $\eta$ compared with MADDPG in the Keep Away task. The shaded region represents the standard deviation of the average evaluation over five trials.}
\label{figure:at}
\end{figure}

\subsubsection{Impact of the Specific Parameter}\label{S4-2-4}

In this subsection, we explored the impact of the special parameter $\eta$ that controls the strength of the relative entropy term in MACDPP.
The learning curves of the Keep Away task with different values of $\eta$ using MACDPP are shown in Fig.~\ref{figure:at}.
With a wide range of $\eta$ from $0.001$ to $0.5$, MACDPP consistently outperformed the baseline method MADDPG in both the mean and standard deviation of the average returns.
It is also observed that a proper selection of $\eta$ could significantly improve MACDPP's learning performance.
The most superior learning curve was obtained when $\eta=0.1$.
An over-small parameter $\eta=0.001$ resulted in extremely slow convergence and a very large standard deviation of return at the beginning.
As a comparison, the large one has less effect on smooth policy updates and may fail in learning more optimal control policies within a limited number of interactions.

\subsection{Cooperation in Traditional Control Task}\label{S4-3}

\begin{figure*}
\centering
\includegraphics[width=1.5\columnwidth]{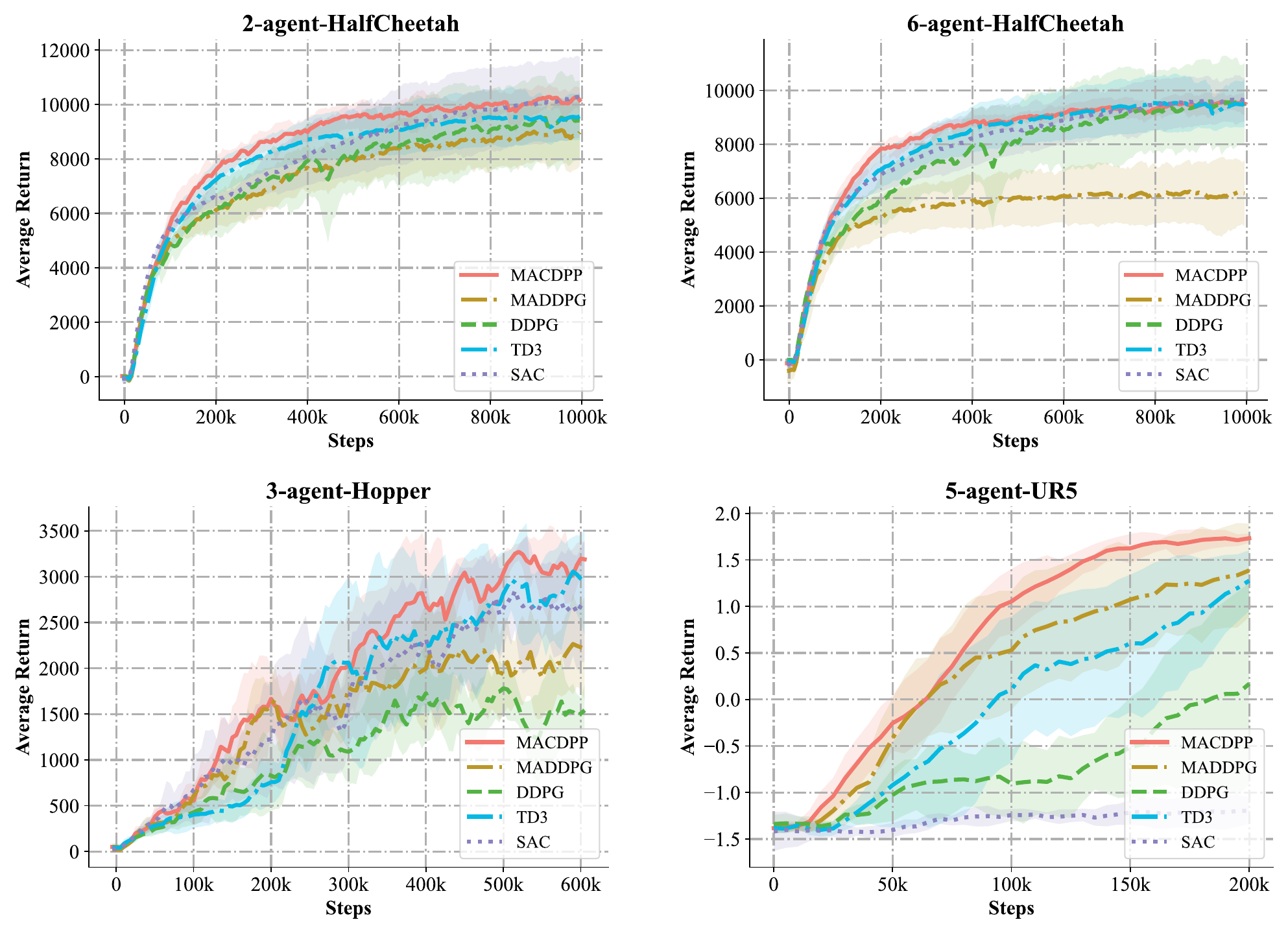}
\caption{Learning curves of MACDPP and all baselines in Mujoco and robo-gym benchmark tasks. The shaded region represents the corresponding standard deviation over five trials.}
\label{figure:lr2}
\end{figure*}

\begin{figure}
\centering
\includegraphics[width=0.85\columnwidth]{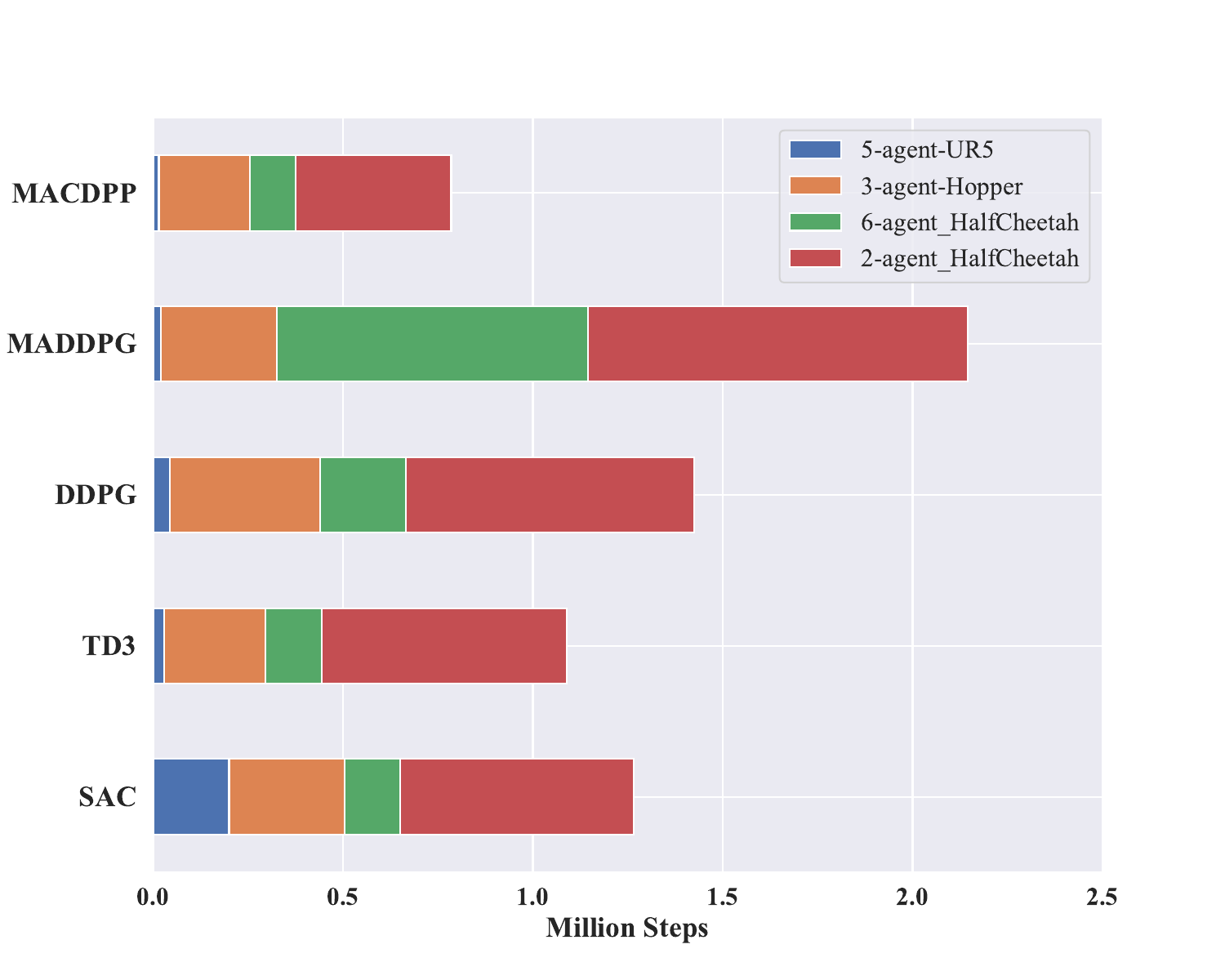}
\caption{Number of interactions utilized by all compared approaches in Mujoco and robo-gym benchmark tasks to reach the lower boundary of maximum average return.}
\label{figure:se2}
\end{figure}

\subsubsection{Evaluation of the Learning Capability}\label{S4-3-1}
In this section, we moved to the traditional control scenarios where the MARL approaches were employed to jointly control one system. In this section, we compared two MARL approaches MACDPP and MADDPG which were treated as the proposed method without using relative entropy regularization. Three widely implemented single-agent RL approaches DDPG, TD3 and SAC were also compared.
The learning curves of all compared approaches in 2-agent HalfCheetah (the system was jointly controlled by MARL with two agents), 6-agent HalfCheetah, 3-agent Hopper and 5-agent UR5 robot arm were illustrated in Fig.~\ref{figure:lr2}.
The average maximum returns in the evaluation phase are listed in Table.~\ref{table:lr2}. Please note that the results of single-agent methods in two HalfCheetah were slightly different since we used difficult random seeds for each task.

In the 2-agent HalfCheetah task, MADPPG learned the worst policy with the lowest average return. In this case, the joint control strategy failed to suppress the single-agent approach while introducing an additional computational burden.
As a comparison, our method converged to the best average return overall compared approaches with a significant advantage in convergence velocity.
Regarding the maximum average return in evaluation, MACDPP outperformed $12.61\%$, $8.57\%$ and $6.0\%$ compared with MADDPG, TD3 and DDPG while achieving slightly better results than SAC with significantly superior converge velocity.
In the 6-agent HalfCheetah task, the learning capability of MADPPG hugely deteriorated so that the six joints could not effectively cooperate.
As a comparison, the proposed method successfully learned the task as well as other single-agent baselines with not only a superior converge velocity but also less standard deviation in the average returns. It enjoyed over $50\%$ more maximum average returns than MADDPG which did not employ the relative entropy regularization. 
In the 3-agent Hopper task, MADDPG outperformed DDPG in average return and converge velocity while the proposed MACDPP achieved overall superior performances than all compared baselines. It quickly converged to the best maximum average return which was $56.07\%$, $25.95\%$, $3.84\%$ and $2.03\%$ than DDPG, MADDPG, SAC and TD3, respectively.
In the more practical UR5 control scenarios where five independent joints were jointly controlled by five agents in MARL methods, both MACDPP and MADDPG outperformed single-agent RL approaches. Within $200$k steps, SAC and DDPG could not learn the task (SAC could not converge at all, and DDPG learned extremely slowly at the first $70$k steps) while only TD3 achieved a close average return to MADDPG.
As a comparison, MACDPP consistently demonstrated superiority in both learning capability and converge velocity, it quickly reached an average return over $0$ within $70$k steps and finally obtained $210\%$, $12.05\%$, $264.6\%$ and $16.3\%$ higher maximum average return than DDPG, MADDPG, SAC and TD3.

\subsubsection{Evaluation of the Sample Efficiency}\label{S4-3-2}

\begin{table*}[h]
	\caption{Average calculation time of $1000$ steps over four benchmark control tasks}
	\centering
	\label{table:time2}
        \resizebox{0.95\linewidth}{!}{
	\begin{tabular}{|l|c|c|c|c|c|}
        \hline
		&\textbf{MACDPP}&\textbf{MADDPG}&\textbf{DDPG}&\textbf{TD3} &\textbf{SAC}\\\hline
		HalfCheetah (2 agents) & $13.84\pm0.29$ s & $9.44\pm0.30$ s  & $4.51\pm0.20$ s & $4.18\pm0.07$ s & $8.36\pm0.10$ s\\\hline
		HalfCheetah (6 agents) & $54.84\pm0.15$ s & $38.35\pm0.29$ s  & $4.66\pm0.24$ s & $4.34\pm0.29$ s & $8.63\pm0.41$ s\\\hline
		Hopper (3 agents) & $22.61\pm0.89$ s & $15.83\pm0.72$ s  & $4.86\pm0.36$ s & $4.69\pm0.32$ s & $8.84\pm0.64$ s\\\hline
	    UR5 (5 agents) & $284.58\pm28.80$ s & $219.32\pm29.31$ s  & $162.03\pm16.44$ s & $177.65\pm22.73$ s & $144.68\pm0.64$ s\\\hline
	\end{tabular}}
\end{table*}

\begin{figure*}
\centering
\includegraphics[width=1.8\columnwidth]{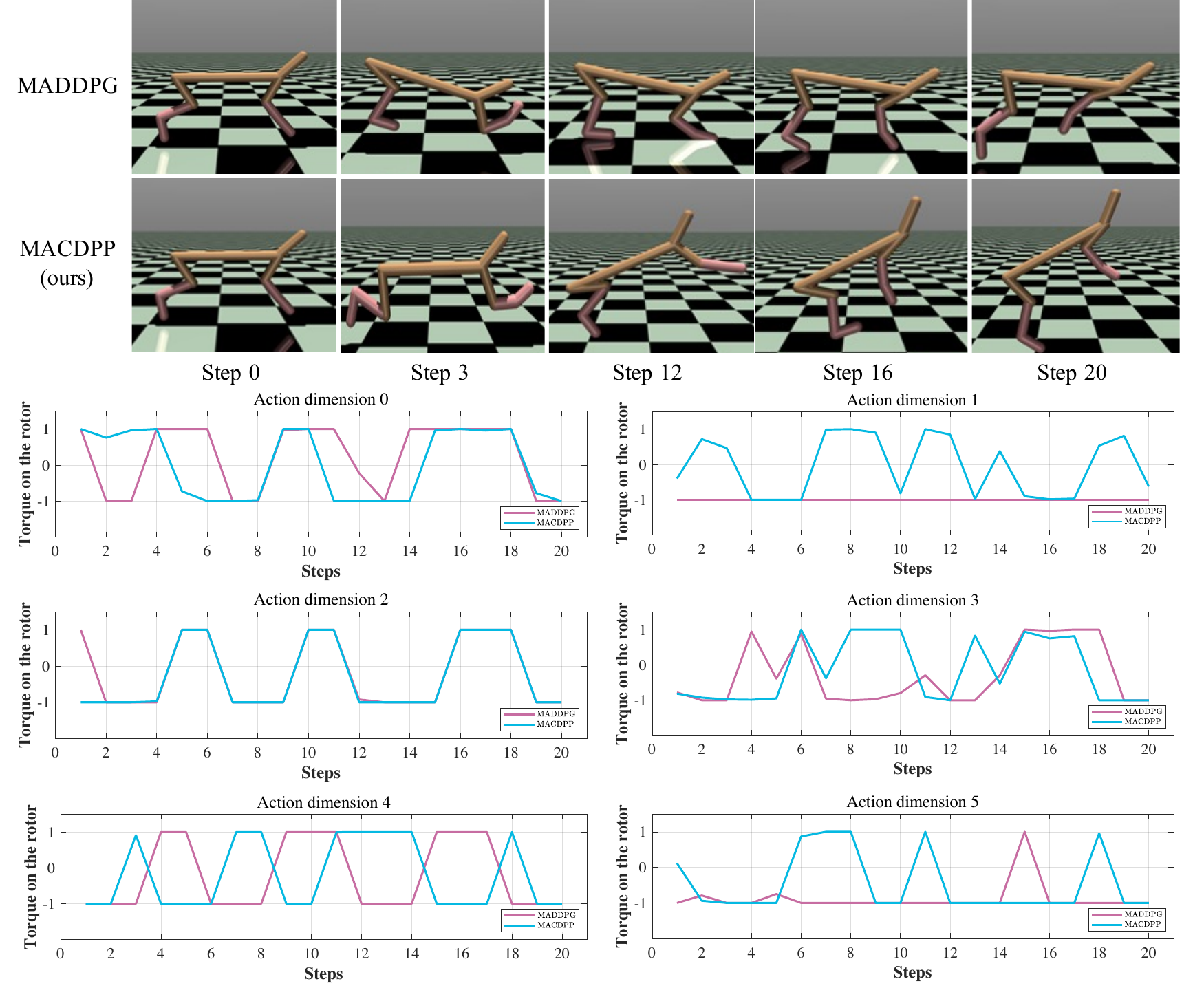}
\caption{Trajectories of control actions using MADDPG and MACDPP in one test rollout of 6-agent HalfCheetah task. The actions of MADDPG and MACDPP were drawn in purple and blue respectively.}
\label{figure:case1}
\end{figure*}

\begin{figure*}
\centering
\includegraphics[width=1.8\columnwidth]{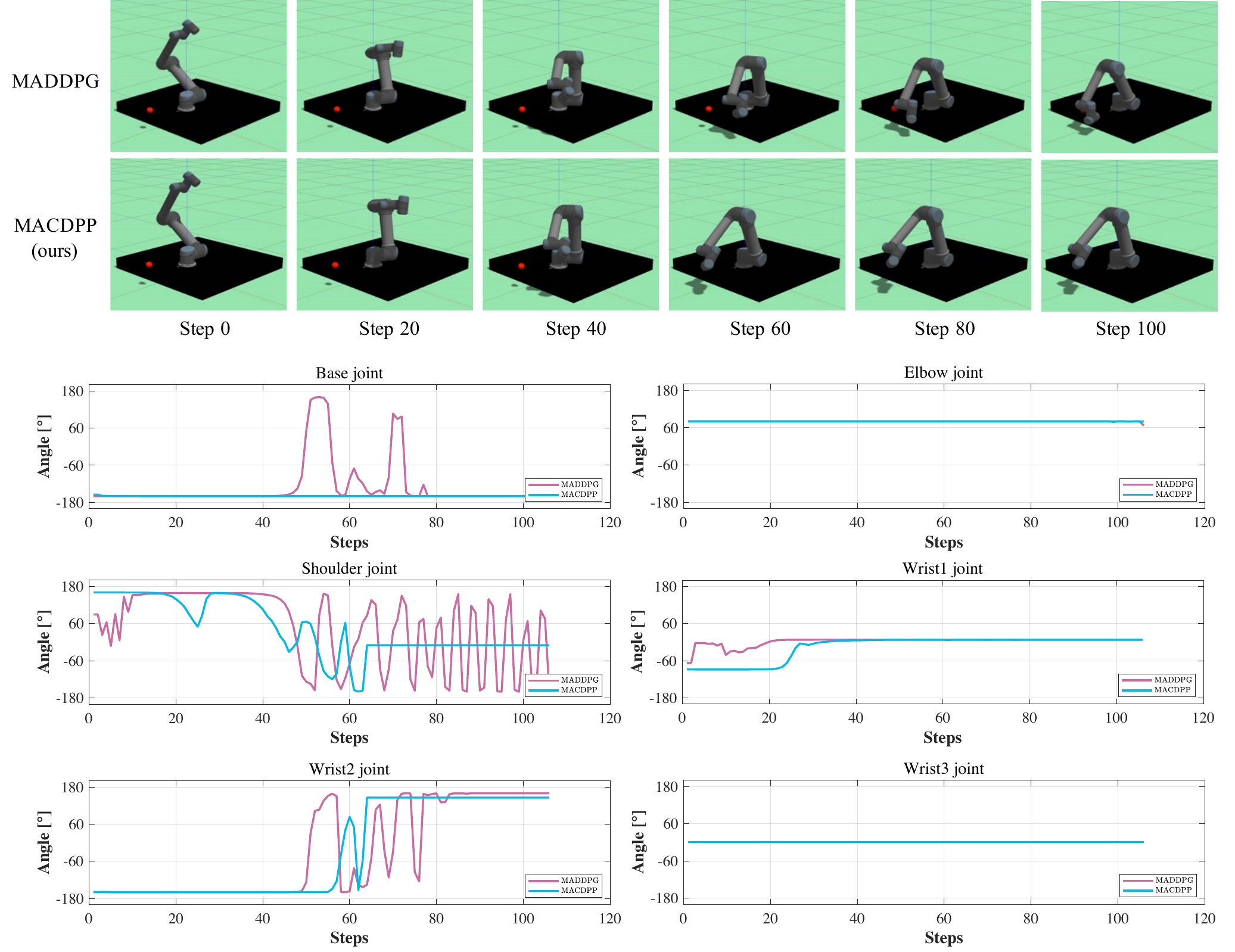}
\caption{Trajectories of control actions using MADDPG and MACDPP in one test rollout of 5-agent UR5 control task ur\_ee\_position . The actions of MADDPG and MACDPP were drawn in purple and blue respectively. The last dimension Wrist3 joint was not included in the controllable action as it was fixed to $0$ in the ur\_ee\_position task.}
\label{figure:case2}
\end{figure*}

The sample efficiency of MACDPP was evaluated in Fig.~\ref{figure:se2}.
Compared with MADDPG which converged slower to a certain level of control performances than the signal-agent baselines, the proposed method demonstrated great advantage in sample efficiency with the regularization of relative entropy. It only spent $36.6\%$ samples to reach the same performance. This result indicated the importance of properly restricting large policy updates in MARL for superior effectiveness.
MCDPP successfully reduced $44.91\%$, $37.94\%$ and $27.98\%$ usage of interactions than DDPG, SAC and TD3.

\subsubsection{Evaluation of the Computational Efficiency}\label{S4-3-3}

The computational burden was evaluated in Table~\ref{table:time2} by summarizing the average calculation time of the first $1000$ steps in four control scenarios.
Although the proposed method required $132.85\%$, $213.5\%\%$, $196.92\%$ and $220.43\%$ more computational times compared with MADDPG, DDPG, TD3 and SAC.
It was observed that the computational burden of MACDPP was alleviated in traditional control scenarios where the system operation took more time. The proposed method additionally consumed $32.85\%$ time than MADDPG.
Compared with the faster single-agent methods, our method increased the computation time from $96\%$ to $120\%$ while employing about four times more agents.
Furthermore, with the increasing system operation and communication times (i.e., from Mujoco to robo-gym based on ROS toolkit), the phenomenon above became more and more noticeable.
It demonstrated the potential and effectiveness of MACDPP in jointly controlling large-scale systems.

\subsubsection{Case Study}\label{S4-3-4}

In this subsection, we investigated the superior control behaviors of MACDPP through the rollouts of learned policies in both 6-agent HalfCheetah and 5-agent UR5 control scenarios.
In the first case study, we explored the learned policies of MADDPG and MACDPP under the same setting of parameter and random seed. The test rollouts with $20$ steps and the corresponding trajectories of each joint were analyzed in Fig.~\ref{figure:case1}.
It is clearly observed that the learned control behavior of MACDPP is more effective than the one of MADDPG.
Effectively coordinating six joints by six agents, the proposed method learned a superior control strategy. Each joint timely conducted proper torque according to the current system states, resulting in a faster movement.
In comparison, MADDPG had significant disadvantages in terms of coordinating six joints by separate agents.
Although the agent in charge of dimension 2 successfully learned a similar policy to the one of MACDPP, the whole multi-agent system struggled to generate proper torques from other agents: the agent of dimension 5 only produced effective torque near the $15$-th step while the agent of dimension 1 was fixed with $-1$ torque during the how rollout.
Due to the lack of relative entropy regularization from the algorithmic perspective, the multiple agents in MADDPG were unable to learn effective and cooperative control strategies.

In the next case study, we studied the test rollouts using MADDPG and MACDPP in the ur\_ee\_position task which aims to control the end-effector of the UR5 robot arm to reach the randomly generated targets.
It is observed that MACDPP quickly drove the robot to finish the task within $60$ steps. The action trajectories showed proper cooperation between each joint.
The base and elbow joints continuously output $-180^{\circ}$ and $60^{\circ}$ throughout the task.
At step $25$, the shoulder joint and wrist 1 joint were coordinated to guide the end-effector to move forward to the target position.
Around step $60$, the shoulder and wrist 2 joints worked together to quickly reach the target. The action Trajectories of all dimensions were effective with minimal jitter.
Compared with our method, MADDPG could not sufficiently learn the cooperative strategy over five joints. The base joint could not continuously output a certain degree, it had strong trembling between steps $40$ to $80$. The shoulder and wrist 1 joints failed to cooperate effectively at the beginning, resulting in redundant movements of the end-effector.
After step $40$, the shoulder and wrist 2 joints could not achieve seamless coordination. Both of them experienced sustained tremors which ultimately resulted in highly degraded control performance.

The experimental results above revealed the advantages of MACDPP in the joint control of robot systems.
The multiple agents reduced the exploration complexity in one system and resulted in faster policy convergence compared to single-agent approaches with the same amount of interactions.
At the same time, the relative entropy regularization significantly avoided the mismatch between the update of multi-agent policies during the learning, promoting the effectiveness in the learning of cooperative control strategies.

\section{Conclusions}\label{S5}
This article proposed a novel MARL approach MACDPP to improve the learning capability and sample effectively in a wide range of control scenarios including multiple agents cooperative/competitive tasks and joint control of a single complicated system. 
It naturally alleviated the inherent inconsistency over multiple agents policy updates by integrating the relative entropy regularization to the AC structure and CTDE framework. 
MACDPP successfully extended FKDPP which has been successfully implemented in the real-world chemical plant by Yokogawa~\cite{8560593,zhu2020scalable} towards a modern approach that supports deep neural networks, AC structure and CTDE framework in order to fit a wider range of control scenarios.
Through evaluation of different benchmark tasks, ranging from multi-agent cooperation/competition to Mujoco simulator and robot arm manipulation, our proposed method consistently demonstrated significant superiority in both learning capability and sample efficiency compared with related multi-agent and single-agent RL baselines.
All these results indicated the potential of relative entropy regularized MARL in effectively learning complex systems divided into multiple agents with lower sampling costs and better control performance.

\bibliographystyle{ieeetr}
\bibliography{paper}

%








\end{document}